\documentclass[aip,apl,preprint,showpacs,10pt,citeautoscript]{revtex4-2}
\fontfamily{cmss}
\usepackage{amsfonts,amsmath,amssymb,epsfig,color,wasysym}
\usepackage{float}
\usepackage{graphicx}%
\usepackage{dcolumn}
\usepackage{bm}
\usepackage{booktabs}
\usepackage{dcolumn}
\usepackage{hyperref}% add hypertext capabilities
\hypersetup{
    colorlinks,%
    citecolor=blue,%
    linkcolor=blue,%
    urlcolor=blue
}

\usepackage{multirow}
\providecommand{\U}[1]{\protect\rule{.1in}{.1in}}

\date{\today}

\begin{document}

\title{\textcolor{black}{Self-compensation by silicon $DX$ centers in ultrawide-bandgap nitrides}}
\author{John L. Lyons}\thanks{john.l.lyons27.civ@us.navy.mil}
\author{Darshana Wickramaratne}\thanks{darshana.k.wickramaratne.civ@us.navy.mil}
\affiliation{Center for Computational Materials Science, US Naval Research Laboratory, Washington,
D.C. 20375, USA}

\begin{abstract}
\textit{DX} behavior limits $n$-type carrier concentrations in ultrawide-bandgap nitrides such as aluminum nitride (AlN) and cubic boron nitride ($c$-BN). Instead of acting as effective-mass donors, \textit{DX} centers capture two electrons, stabilizing a negative charge state that leads to self compensation. Silicon is the most effective $n$-type dopant in this class of materials; in AlN, its \textit{DX} level [(i.e., the (+/$-$) transition level] is $\sim$270 meV from the conduction-band minimum. This implies that many silicon impurities incorporated into AlN will be negatively charged and compensate the intended $n$-type doping. By combining density functional theory calculations of temperature-dependent band gaps and Si dopant transition levels, we show here that significant compensation occurs in silicon-doped AlN, even in the absence of any other defects. This compensation strongly limits free electron concentrations which become independent of doping concentration, and donor activation is only significant for light doping scenarios. Higher free carrier concentrations can be achieved in AlGaN alloys or in $c$-BN, where the \textit{DX} level sits closer to the conduction-band minimum.
\end{abstract}

\maketitle

\section{Introduction}
\label{intro}
Ultrawide-bandgap materials, which are semiconductors with bandgaps in excess of 3.5 eV, are promising for a variety of applications including high-frequency and high-power devices, as well as devices for extreme
environments.\cite{Tsao18,Xu22} In particular, ultrawide-bandgap (UWBG) nitride semiconductors appear to be among the best candidates for power electronics and radio frequency technologies.\cite{Kaplar17,Woo24} Nitride-based materials are also leading candidates for devices operating at high temperatures.\cite{Liu2026} However, controlling electrical conductivity in UWBG semiconductors has proven even more challenging than in wide-bandgap semiconductors such as gallium nitride \cite{Tsao18,Lyons24}.

As is the case for many UWBG materials, hole localization at acceptor impurities leads to large ionization energies, presenting a major challenge for achieving $p$-type conductivity.\cite{Tsao18,Lyons24} In contrast, efforts to achieve $n$-type conductivity in UWBG nitride semiconductors have seen moderate success. For example, doping AlN and $c$-BN with silicon was reported to lead to free electron concentrations {between $\sim$10$^{14}$ and $\sim$10$^{15}$ cm$^{-3}$} at room temperature \cite{Zeisel00,Taniyasu04,Taniyasu06,Thapa08,Contreras16,Bagheri22,Mukhopadhyay24}. Si implantation\cite{Breckenridge21} and Si doping of N-polar AlN Si \cite{Khan24} have been reported to lead to even higher carrier concentrations. However, the non-monotonic dependence of the carrier mobility and conductivity versus both Si concentration and annealing conditions makes it challenging to deduce the ionization energy of Si in AlN in these studies.

\begin{figure}
\centering
\includegraphics[width=0.6\textwidth]{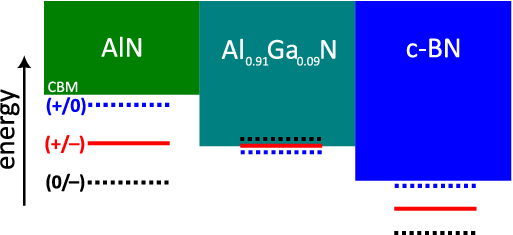}
\caption{Schematic of the thermodynamic transition levels calculated for Si \textit{DX} centers in AlN,\cite{Wang26} Al$_{\rm 0.91}$Ga$_{\rm 0.09}$N, and $c$-BN\cite{Turiansky21}.
}\label{levels}
\end{figure}

A hydrogenic model for dopant ionization energies suggests that both of these UWBG nitrides should exhibit low donor ionization energies. For AlN, which has an effective electron mass of 0.33 $m_0$ and a dielectric constant of 9.14, effective-mass theory predicts a donor binding energy of 54 meV.\cite{Ashcroft1976} For $c$-BN using a density of states effective mass of 0.43 $m_0$ \cite{Hirama20} and a dielectric constant of 3.9 leads to a donor binding energy of 38 meV. While these estimates are independent of donor species, they indicate that if $n$-type doped, donors should be thermally ionized at room temperature. First-principles calculations, which can directly interrogate the electronic structure of a given dopant, have been used to consider the prospects for $n$-type doping of both compounds. Candidate substitutional dopants were found to either lead to \textit{DX} levels within the band gap of the material (as in the case of Si, O, and Ge donors),\cite{Gordon14,Wang26,Turiansky21,Liu25,Ganose26} or have deep (but non-\textit{DX}) donor levels that will not lead to thermal ionization of electrons (as in the case of the chalcogen donors).\cite{Lyons25} {(Recently Al$_i$ was predicted to be a shallow donor in AlN,\cite{Liu25} but only in metastable configurations.)}

When a donor exhibits \textit{DX} behavior, self compensation can occur via the formation of an acceptor state (created by the capture of two electrons), in a process defined in Ref.~\onlinecite{Park97} as:
\begin{eqnarray}
\label{eq:dx}
2d^0 \rightarrow d^+ + \textit{DX}^-,
\end{eqnarray}
where $d^0$ is the neutral donor, $d^+$ is the positively charged donor dopant, and \textit{DX}$^-$ represents the dopant after capturing two electrons and becoming negatively charged. Equilibrium between the positive and
negative charge states of the dopant can be defined by the resulting thermodynamic transition level between these charge states [written as (+/$-$)], which is given as:
\begin{eqnarray}
\label{eq:pm}
\varepsilon (+/-) = \frac{E^f(d^+) - E^f(\textit{DX}^{-})}{2},
\end{eqnarray}
where $E^f(d^+)$ [$E^f(DX^-)$] is the formation energy of $d^+$ [$\textit{DX}^{-}$] when the Fermi level is referenced to the same energy, such as the conduction-band minimum (CBM). Defect formation energies can be calculated using first-principles methods as described in Ref.~\onlinecite{RMP}, and the other thermodynamic transition levels of the Si dopant [namely, (0/+) and (0/$-$)] can be written using expressions analogous to Eq.~\ref{eq:pm}. Equilibrium between these two charged species, which is established by the position of the (+/$-$) thermodynamic transition level, is crucial for understanding the behavior of \textit{DX} centers. As we will see, it roughly determines the limit for how close the Fermi level can approach the CBM. Thus, the further the (+/$-$) level resides from the CBM, the fewer free electrons will be generated by the given dopant.  Among the donor candidates, silicon substituting on the cation site has been found to lead to a \textit{DX} level that is closest to the conduction band.\cite{Gordon14}
\par
Some studies have attributed the limited carrier concentrations reported in Si-doped AlN and $c$-BN to compensation by acceptor defects (either native point defects or impurities) \cite{Chichibu13,Harris18,Ganose26}.
In this scenario, an acceptor defect is incorporated at nominally the same concentration as Si, and the resulting Fermi level will be pinned near the intersection of the Si donor formation energy and the lowest-energy acceptor. However, the \textit{DX} center itself can also be a major source of compensation\cite{Boguslawski97,vdwDX98,Mccluskey98,Zeisel00,Son11,Gordon14,Mehnke16,Prozheev25,Wang25}. Distinguishing between the two scenarios is important, since compensation by the formation of \textit{DX} centers is inherent to the dopant, whereas extrinsic compensation could be overcome by reducing the concentration of the compensating species.
\par
Achieving more efficient $n$-type doping would be a great benefit for UWBG nitride-based devices. Because of the compensation that occurs, higher amounts of doping have been explored for AlN. Unfortunately, higher concentrations of \textit{DX} centers are predicted to lead to a collapse in mobility\cite{Wang25} due to the presence of more charged species, an effect that is also observed in experiments.\cite{Taniyasu06} Because of these limitations, some groups have employed polarization-doping strategies\cite{Simon10,Kumabe24,Hiroki25} to enhance conductivity, which if successful could obviate the need for any impurity dopant.
\par
First-principles calculations have predicted that the (+/$-$) transition level of Si$_{\rm Al}$ in AlN is 271 meV below the CBM.\cite{Wang26} [Though a value of 150 meV was reported in an earlier study (Ref.~\onlinecite{Gordon14}), the lower-energy Si$_{\rm Al}^-$ configuration reported in Ref.~\onlinecite{Wang26} pushes the (+/$-$) level further from the CBM; our calculations confirm the findings of Ref.~\onlinecite{Wang26}.] Hence, Si$_{\rm Al}^0$ is never a stable charge state within the band gap. Thus, as shown in Fig.~\ref{levels}, the (0/$-$) thermodynamic transition level is furthest from the AlN CBM, and the (+/0) level is the closest. Furthermore, as discussed in Ref.~\onlinecite{Gordon14}, the (+/$-$) level should move closer to the CBM in AlGaN alloys with increasing Ga content. Specifically, assuming an AlN/GaN valence-band offset of 0.25 eV (with the VBM of AlN being lower in absolute energy)\cite{Jin20}, and a bowing parameter of 0.7 eV,\cite{Vurgaftman03} we predict that the \textit{DX} level of Si will coincide with the CBM in an Al$_{\rm 0.91}$Ga$_{\rm 0.01}$N alloy. In $c$-BN, silicon is also predicted to be a \textit{DX} center, but with a (+/$-$) level only 110 meV below the CBM.
\par
At 0 K, AlN has a band gap of 6.2 eV, and experimentally reported band gaps of $c$-BN range from 6.1 eV to 6.4 eV. However, devices often operate at elevated temperatures which can lead to a significant reduction in the band gap relative to their 0 K values. This also implies a change in the position of the CBM, and also a potential change in position of the (+/$-$) transition level with respect to the CBM. As discussed below, we account for this temperature dependence by explicitly calculating semiconductor band gaps and band-edge positions as a function of temperature.
\par
Together with this information, we explore how the \textit{DX} behavior of Si dopants leads to self compensation in AlN, the Al$_{\rm 0.91}$Ga$_{\rm 0.09}$N alloy, and $c$-BN. We calculate concentrations of free electrons and charged dopant species as a function of temperature and silicon dopant concentration, using results from prior first-principles calculations as input and incorporating the effects of temperature on the band gap. We show that, considering only compensation by \textit{DX} centers, room-temperature free carrier concentrations are strictly limited (to near 3$\times$10$^{14}$ cm$^{-3}$) in Si-doped AlN, even with Si concentrations as high as 10$^{20}$ cm$^{-3}$. {Alloying Ga into AlN brings the Si (+/$-$) level closer to the alloy CBM, increasing room-temperature free electron concentrations, which then show an increase with increasing Si concentrations}. In comparison with AlN, Si donors in $c$-BN have a (+/$-$) level closer to the CBM, and $n$-type doping of this material is more feasible, with potentially high carrier concentrations accessible at room temperature if external compensation can be avoided.

\section{Methodology}
\begin{table}[]
\begin{center}
\centering
\caption{First-principles calculations of transition levels of Si donors in AlN, AlGaN, and $c$-BN, as reported in Refs.~\onlinecite{Wang26,Turiansky21}.}
\setlength{\tabcolsep}{6pt} % Default value: 6pt
\renewcommand{\arraystretch}{1.4} % Default value: 1
\begin{tabular}{l|ccc}
  \toprule
  {compound} & {(+/0)} & {(+/$-$)} & {(0/$-$)} \\
  \midrule
  AlN & 50 & 271 & 492 \\
  Al$_{\rm 0.91}$Ga$_{\rm 0.09}$N & 50 & 0 & $-$50 \\
  $c$-BN & 10 & 110 & 210 \\
   \bottomrule
\end{tabular}
\label{tab:ttls}
\end{center}
\end{table}
\par
We use as input the first-principles calculations of donor transition levels from Ref.~\onlinecite{Wang26} for Si$_{\rm Al}$ in AlN and Al$_{\rm x}$Ga$_{\rm 1-x}$N, and from Ref.~\onlinecite{Turiansky21} for Si$_{\rm B}$ dopants in $c$-BN. A schematic of these levels are shown in Fig.~\ref{levels}, and Table~\ref{tab:ttls} lists the explicit values we use to solve for charge neutrality and determine the concentrations of charged species at given temperatures and dopant concentrations.
\par
The transition levels listed in Table \ref{tab:ttls} apply to a band structure at the 0 K limit. However, as temperature increases, the band gaps and band-edge positions of semiconductors change.  We incorporate these effects by calculating the temperature dependence of band edges with explicit density functional theory (DFT) calculations.
To calculate band gap renormalization of AlN and $c$-BN we use DFT and the one-shot method \cite{zacharias2015stochastic,zacharias2016one} as implemented in VASP \cite{karsai2018electron} with the r$^{2}$SCAN meta-GGA functional \cite{furness2020accurate}.  While the r$^{2}$SCAN meta-GGA functional underestimates the absolute band gaps of AlN and $c$-BN, we are interested in the relative change in band gaps as a function of temperature, which we expect to be adequately described given that r$^{2}$SCAN has been shown to accurately describe the vibrational properties of materials\cite{ning2022reliable}.  All of the temperature dependent calculations employ a 6$\times$6$\times$6 supercell, which we verified by comparing results at 800 K obtained within an 8$\times$8$\times$8 supercell.  The band gaps of AlN and $c$-BN in these two supercells are converged to within 10 meV. We use an energy convergence threshold of 10$^{-7}$ eV for all calculations. We calculate the temperature dependence of the band gap, using the equilibrium static lattice constants obtained from optimizing the structures with DFT. {For the quasi-harmonic approximation (QHA) calculations, we use the thermal expansion coefficients ($a$=5.3$\times$10$^{-6}$ K$^{-1}$ and $c$=4.2$\times$10$^{-6}$ K$^{-1}$ for AlN \cite{yim1974thermal} and $a$=3.3$\times$10$^{-6}$ K$^{-1}$ for $c$-BN \cite{datchi2007equation}) applied to the $T$=0K lattice constants.}

In AlN, the valence-band maximum (VBM) originates from the split-off band, which is well separated from the heavy hole bands. We therefore define the renormalized band gap using the highest occupied and lowest unoccupied state energies from the supercell calculations at each temperature. In $c$-BN the supercell calculations fold the three equivalent $X$ valleys to $\Gamma$. While these zone-folded conduction band states are degenerate at 0 K, atomic displacements at finite temperature lift the degeneracy with separations of $\sim$150 meV. We define the CBM using the lowest conduction state and the VBM as the average of the top three valence states.

The temperature dependence of the band edges within the QHA is determined by a combination of electron-phonon interaction and thermal expansion. The temperature dependence of the conduction band minimum due to electron-phonon interaction is obtained from our static-lattice calculations ($T$=0 K lattice parameters), where the {eigenvalues} for each supercell calculation are referenced to the same electrostatic potential.  To determine the contribution due to thermal expansion, we used the thermally induced change in volume at each temperature and the absolute deformation potential of the band edges \cite{van1997small} to determine the resulting shift in the band edges.  The conduction band deformation potential of $c$-BN is determined by calculating the band gap deformation potential due to hydrostatic strain and using the valence band deformation potential of AlN, \cite{van1997small} since the valence band maximum of $c$-BN is also derived from N $2p$ states. We verified that the combination of this shift in band edges due to thermal expansion and the shift in band edges due to electron-phonon interaction (obtained at the static-lattice level of theory using the $T$=0 K lattice parameters) leads to a change in band gap that agrees with our QHA calculations of the change in the band gap.
\par
Note that we neglect the temperature dependence of the \textit{DX} transition levels themselves. This is based on our observation that for the temperature range that we consider in this study, deep defect levels exhibit a weaker shift with temperature compared to band-edge shifts \cite{wickramaratne2018defect}.
Since the Si (+/0) donor levels are effective-mass-like, deriving their character from perturbed conduction-band states, we anticipate that these levels will track the conduction band with temperature. Thus, we leave their ionization energies fixed with respect to the CBM for all temperatures.
We solve for the charge neutrality equation to determine defect and free carrier concentrations as well as the Fermi level by using the SC-FERMI\cite{Buckeridge19} code. These calculations require energy-dependent density of states from the bulk material; these were obtained from HSE DFT calculations as described above. For the charge-neutrality calculations, we set varying overall concentrations of Si$_{\rm Al}$ and then solve for charge neutrality, allowing the concentrations of electrons, holes, Si$_{\rm Al}^+$, {Si$_{\rm Al}^0$,} and Si$_{\rm Al}^-$ to vary. We assume that Si is the dominant species within the material, and no other defects are explicitly included. This allows us to isolate the effects of \textit{DX} behavior on $n$-type doping. We also neglect the effect of short-range order around Si dopants in the Al$_{\rm 0.91}$Ga$_{\rm 0.09}$N alloy,\cite{Prozheev25,Ganose26} as this is beyond the scope of our study. Although our setup is an idealized model of Si doping in UWBG nitrides, it is intended to represent a ``best-case scenario'' for exploring what level of free electron concentrations can be achieved in these materials.
\par
It should be noted that at high concentrations, the Bohr radii of donors may begin to interact,\cite{Mott61} potentially enabling transport through impurity-band conduction and lowering the apparent ionization energy of the donor. For AlGaN alloys, a transition to Mott behavior has been observed to occur at carrier densities {varying from 3$\times$10$^{18}$ (in Ref.~\onlinecite{Bharadwaj19}) to 3$\times$10$^{19}$ cm$^{-3}$ (in Ref.~\onlinecite{Zhu04}).} Though a fuller discussion of this point is beyond the scope of this work, the free carrier concentrations that we calculate in this work for Si-doped nitrides are below this threshold in the majority of scenarios we consider.

\section{Results and discussion}
\subsection{Temperature dependence of band gap}
\begin{figure}[!htb]
\centering
\includegraphics[width=0.5\textwidth]{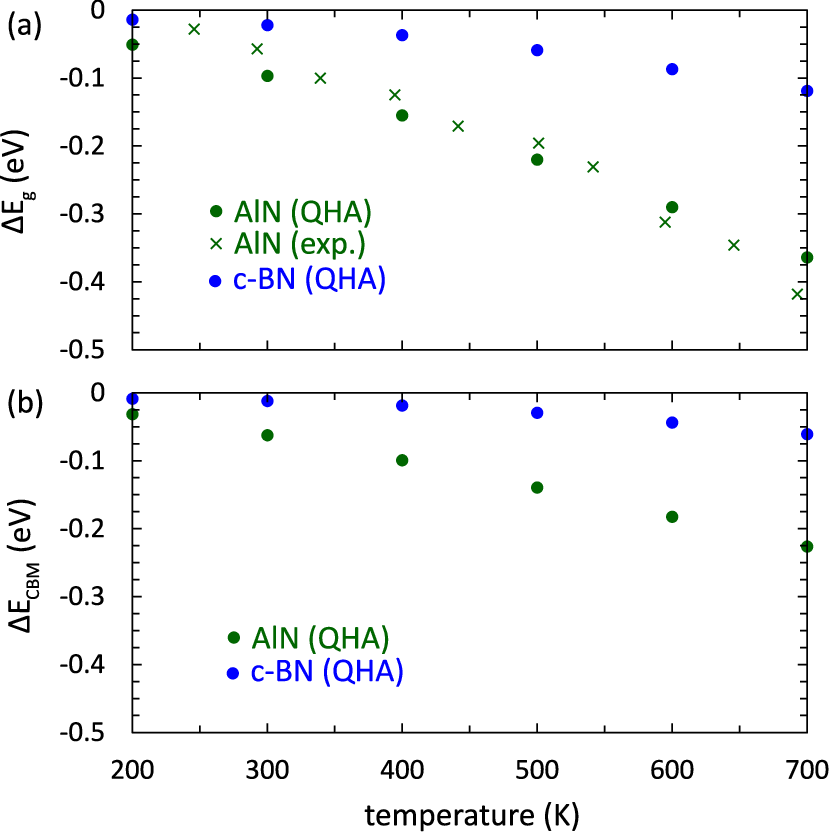}
\caption{(a) Calculated change in the fundamental band gap within the quasi-harmonic approximation (green circles for AlN, and blue circles for $c$-BN) as a function of temperature.  The experimental data (green x's) are obtained from Ref.~\onlinecite{nam2004optical}.
(b) Calculated change in the energy of the conduction band minimum relative to its value at $T$=0 K, $\Delta E_{\rm CBM}$ for AlN and $c$-BN.  The color scheme is similar to that used in panel (a).
}
\label{fig:eg_T}
\end{figure}
\par
We begin by calculating the temperature dependence of the band gap ($E_{g}$) for both materials using DFT. We employ the QHA which accounts for the thermal expansion of the lattice at each temperature [Fig.~\ref{fig:eg_T}(a)]. Using the QHA, we find that the $E_{g}$ of AlN decreases by 364 meV at 700 K, while in $c$-BN $E_{g}$ decreases by 119 meV. Notably, {we find that} the temperature dependence of $E_{g}$ is larger in AlN compared to $c$-BN across the entire temperature range considered here. Comparing our calculated QHA results for the temperature-dependent $E_{g}$ of AlN with optical measurements, we find good agreement with experiment \cite{nam2004optical}. We are not aware of any experimental studies of the temperature dependence of the band gap of $c$-BN, but the weak temperature dependence that we find in our calculations is consistent with the conclusions of Watanabe {\it et al.} who also observed a weak change in the band gap of $c$-BN \cite{watanabe2004ultraviolet}
\begin{figure*}
\centering
\includegraphics[width=\textwidth]{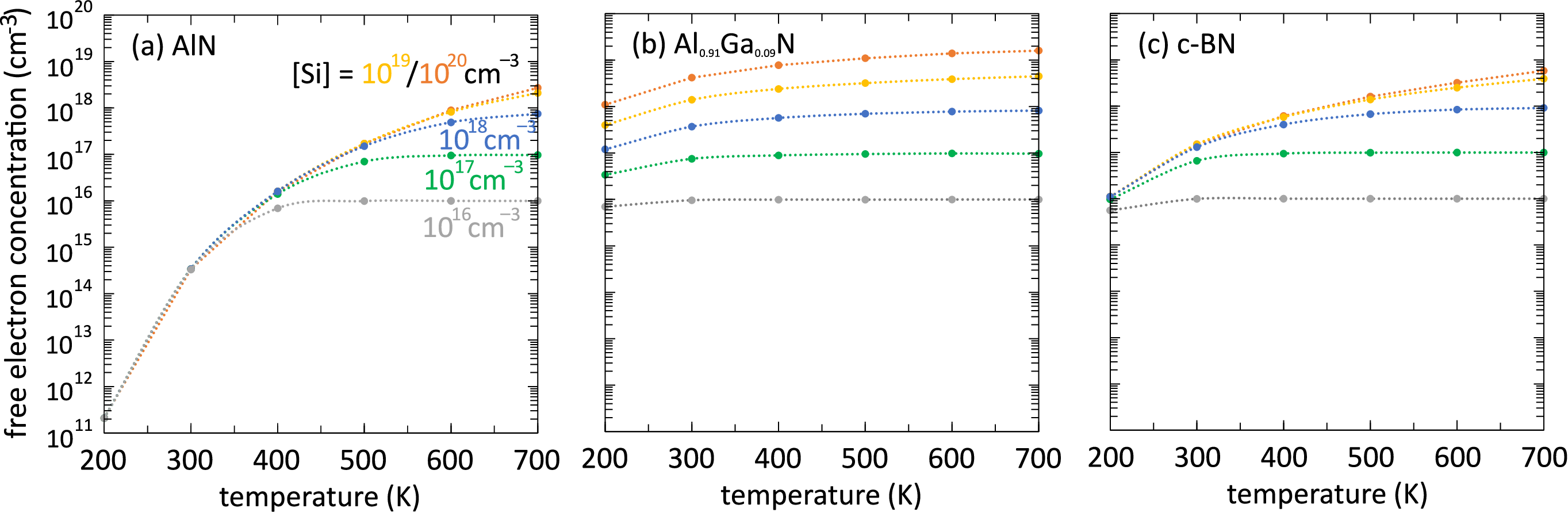}
\caption{Free carrier concentrations (electrons per cm$^{-3}$) vs temperature (K) for (a) AlN, (b) Al$_{\rm 0.91}$Ga$_{\rm 0.09}$N, and (c) $c$-BN containing various concentrations of Si$_{\rm Al}$. Gray dots assume a silicon concentration of 10$^{16}$ cm$^{-3}$, green are for 10$^{17}$ cm$^{-3}$, blue are for 10$^{18}$ cm$^{-3}$, yellow are for 10$^{19}$ cm$^{-3}$, and orange are for 10$^{20}$ cm$^{-3}$. Dotted lines are a guide to the eye.
}\label{cvt}
\end{figure*}
\par
An advantage of our calculations is that we can identify the shifts of individual band edges [Fig.\ref{fig:eg_T}(b)]. Within the QHA, the AlN CBM shifts down in energy by 230 meV at 700 K relative to its energy at 0 K.  In $c$-BN, the QHA calculations predict the CBM shifts down by 64 meV at 700 K relative to the 0 K energy of the CBM. The CBM in AlN is at $\Gamma$, while in $c$-BN it is at the $X$ valley, which may partly explain the difference in quantitative shifts of the band edges in both materials. In the following, we assume that the temperature dependence of the relative band-edge shifts of the Al$_{\rm 0.91}$Ga$_{\rm 0.09}$N alloy are the same as for AlN.
\subsection{Silicon doping of AlN}
Using the temperature-dependent band structures described above, we predict free electron concentrations for Si-doped AlN, which are plotted versus temperature in Fig.~\ref{cvt}a. We consider concentrations of Si dopants varying from 10$^{16}$ cm$^{-3}$ (which we will call ``light doping'') to 10$^{20}$ cm$^{-3}$ (which we will call ``heavy doping''), as indicated by the color coding in the plot, and temperatures varying from 200-700 K. As shown in Fig.~\ref{cvt}a, the free carrier concentrations for all doping levels practically overlap at temperatures lower than 300 K. In all doping scenarios free electron concentrations are near 3.4$\times$10$^{14}$ cm$^{-3}$ at room temperature, and the Fermi level is established 0.23 eV below the CBM of AlN (thus limiting the number of free electrons that can be excited). Increasing Si concentrations in AlN beyond light doping therefore does not lead to an increase in free electron concentrations, at least near room temperature.

The major difference between the different doping scenarios in AlN is the concentration of charged Si dopants (i.e., Si$_{\rm Al}^+$ and Si$_{\rm Al}^-$). For light doping, slightly more than half {(52\%)} of silicon exist as Si$_{\rm Al}^+$, and nearly all of the remaining silicon {(48\%)} are present as Si$_{\rm Al}^-$. The concentration of charged Si species is thus roughly equal to the total amount of Si; only a very small amount (10$^{13}$ cm$^{-3}$) are present as Si$_{\rm Al}^0$. This is also the case for heavy doping at room temperature, where Si$_{\rm Al}^+$ $\approx$ Si$_{\rm Al}^-$ $\approx$ 5$\times$10$^{19}$ cm$^{-3}$. Because the Fermi level is locked near the position of the (+/$-$) level, increasing Si doping only leads to an increase in charged impurities, with no subsequent increase in free electron concentrations.

We can further examine this behavior by plotting donor activation as a function of Si concentration, as is done in Fig.~\ref{act}. For light doping ([Si]=10$^{16}$ cm$^{-3}$), Si donor activation is slightly above {3\%}. But the activation ratio drops steadily as more Si is added to the system; at the highest doping levels donor activation is far lower than {0.001\%.} At higher temperatures there is a modest increase in donor activation; for light doping, nearly 100\% of donors are ionized at 500 K. However, even at these temperatures, the activation ratio also quickly drops as the Si concentration increases. For instance, at 500 K, carrier concentrations are limited to {1.7$\times$10$^{17}$ cm$^{-3}$} even for heavy doping, and donor activation is still far less than {1\%.}

\subsection{Silicon doping of AlGaN}

One strategy to increase doping activation could be to alloy Ga into AlGaN, which causes the alloy CBM to move closer to the Si \textit{DX} level.\cite{Gordon14} As shown in Fig.~\ref{levels}, for the Al$_{\rm 0.91}$Ga$_{\rm 0.09}$N alloy, the (+/$-$) level of the Si donor coincides with the alloy CBM. This decreases the likelihood of Si$_{\rm Al}^-$ acting as a compensating species, allowing for a higher activation of Si donors.

For Al$_{\rm 0.91}$Ga$_{\rm 0.09}$N, assuming light doping at room temperature, nearly all Si donors are activated and uncompensated, and exist as Si$_{\rm Al}^+$, as can be seen in Fig.~\ref{cvt}b. Because almost all Si donors give rise to free electrons at 300 K, free carrier concentrations {for light doping} remain nearly flat as the temperature rises for the lightly doped case.

Moving to higher levels of Si doping increases the free electron concentration in Al$_{\rm 0.91}$Ga$_{\rm 0.09}$N at room temperature, in contrast to the AlN case. For Si doping levels of 10$^{18}$ cm$^{-3}$, {3.7$\times$10$^{17}$ cm$^{-3}$} free electrons are produced, and the Fermi level is only {0.05 eV} below the CBM. Under heavy doping the \textit{DX} character of Si has a strong impact.  Doping with 10$^{20}$ cm$^{-3}$ of Si gives rise to {4.2$\times$10$^{18}$ cm$^{-3}$} free electrons at room temperature, and the Fermi level is established {0.01 eV} above the CBM.

Alloying Ga into AlN might thus provide a means for increasing donor doping efficiency, as shown in Fig.~\ref{act}. Adding 9\% of Ga into AlN yields a nearly {29$\times$} increase in carriers for light doping ([Si] = 10$^{16}$ cm$^{-3}$), and a {1000$\times$} increase in carriers for doping of 10$^{18}$ cm$^{-3}$ silicon. However, alloying Ga into AlN can lead to a drastic decrease in mobility due to random alloy scattering.\cite{Coltrin17} Ref.~\onlinecite{Coltrin17} shows that for Al$_{\rm 0.9}$Ga$_{\rm 0.1}$N alloys, mobilities are less than 20\% of the AlN value. It should be noted that Si doping itself will also lead to a decrease in mobility, even in pure AlN. This is because \textit{DX}-center (Si$_{\rm Al}^-$) compensation of Si donors (Si$_{\rm Al}^+$) leads to large concentrations of ionized impurities.\cite{Wang25} Thus, optimizing conductivity for Si-doped AlGaN will necessarily involve a tradeoff between Ga and Si contents. The predictions above for AlN and AlGaN are consistent with prior measurements\cite{Zeisel00,Nakarmi04,Collazo11,Mehnke16,Maeda22} of Si-doped Al$_{\rm x}$Ga$_{\rm 1-x}$N alloys, which indicate that Si remains an effective donor up to high Al contents and the observation that Si activation energies steadily increase with increasing Al content, as well as reports of limited donor activation and large donor activation energies for Si-doped
AlN.\cite{Zeisel00,Taniyasu04,Harris18,Quinones23}

\subsection{Silicon doping of $c$-BN}
Finally, we consider Si doping of $c$-BN, in which the Si$_{\rm B}$  donor has a (+/$-$)  \textit{DX} level 0.11 eV below the CBM. We find that, as would be expected, the behavior of Si-doped $c$-BN falls between AlN and Al$_{\rm 0.91}$Ga$_{\rm 0.09}$N. Higher electron concentrations can be achieved in $c$-BN than in AlN: for light doping, {30$\times$} more electrons (10$^{16}$ cm$^{-3}$) can be generated at room temperature. Unlike the case for Si-doped AlN, free electron concentrations increase modestly with Si concentration. For Si doping of 10$^{18}$ cm$^{-3}$, {1.3$\times$10$^{17}$ cm$^{-3}$} free electrons are generated at 300 K. However, beyond this Si concentration, free carrier concentrations saturate: for 10$^{20}$ cm$^{-3}$ Si doping, only {1.5$\times$10$^{17}$ cm$^{-3}$} free electrons are present. At higher temperatures, there is some variation in free carrier concentrations for Si-doped $c$-BN. As was the case for Al$_{\rm 0.91}$Ga$_{\rm 0.09}$N, light doping does not exhibit temperature dependence, since activation is already high even at 300 K. However, for the heavily doped case, at 500 K, free electrons increase to 1.6$\times$10$^{18}$ cm$^{-3}$, which is over {10$\times$} the room-temperature concentration.

\begin{figure}
\centering
\includegraphics[width=0.5\textwidth]{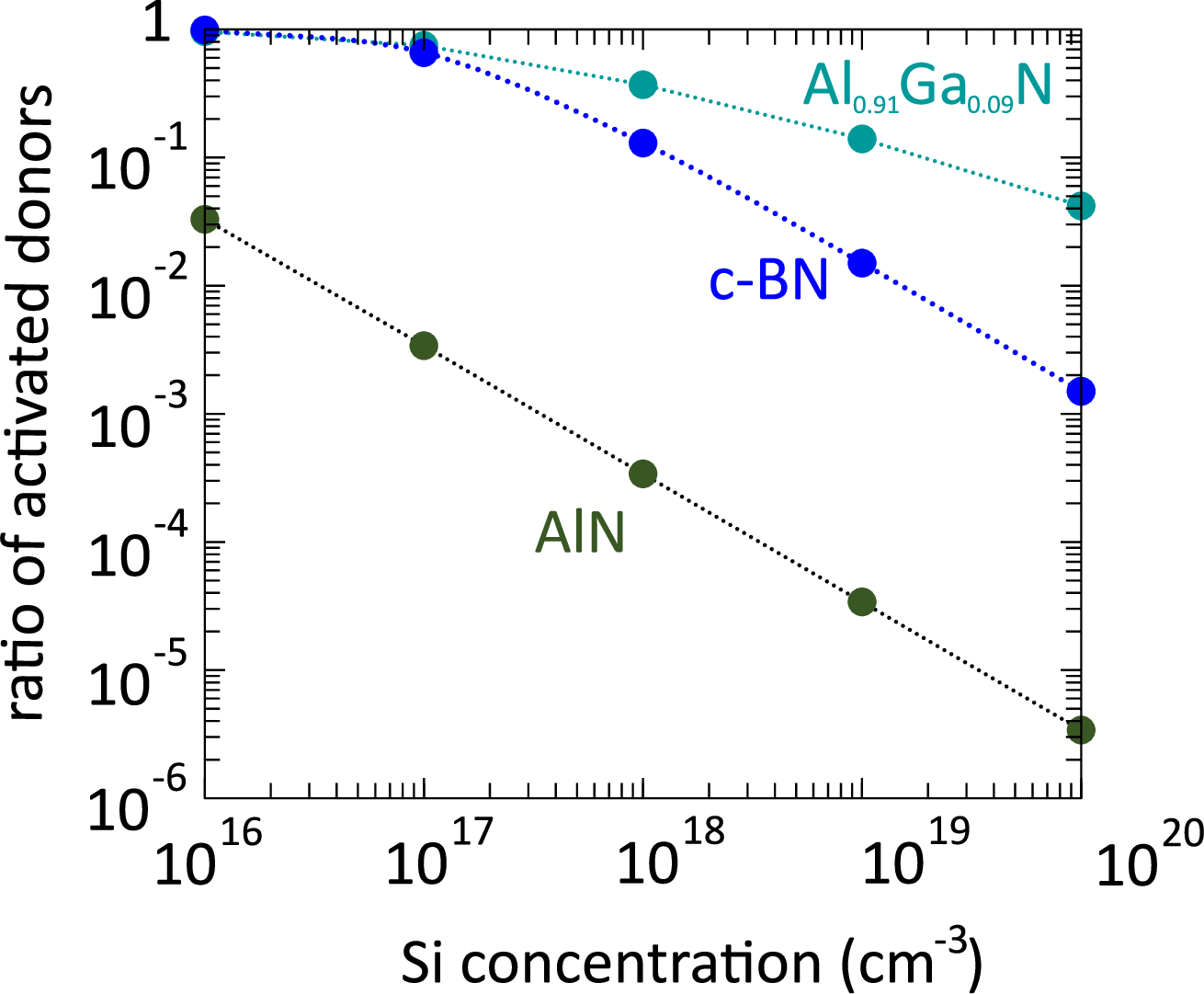}
\caption{Ratio of activated and uncompensated Si donors as a function of Si concentration (in cm$^{-3}$) at room temperature (300 K) for AlN (in dark green), Al$_{\rm 0.91}$Ga$_{\rm 0.09}$N (in teal), and $c$-BN (in turquoise).
}\label{act}
\end{figure}

Thus, as in Al$_{\rm 0.91}$Ga$_{\rm 0.09}$N, for light Si doping of $c$-BN activation ratios are high and compensation is low, as shown in Fig.~\ref{act}. However, at higher Si concentrations, compensation increases and doping efficiency worsens compared to Al$_{\rm 0.91}$Ga$_{\rm 0.09}$N; for heavy doping, less than 1\% of Si are generating free electrons at room temperature. Heavy compensation also occurs for heavy doping, as Si$_{\rm B}^+$ and Si$_{\rm B}^-$ are present in nearly equal concentrations ($\approx$5$\times$10$^{19}$ cm$^{-3}$).

These results show that $n$-type doping may be more feasible in $c$-BN than in AlN. As discussed above, the case of $c$-BN falls between Al$_{\rm 0.91}$Ga$_{\rm 0.09}$N and AlN, and light $n$-type conductivity appears to be feasible. In fact, the free electron concentrations predicted here are actually higher than what have been observed in Ref.~\onlinecite{Hirama20}, possibly due to the presence of compensating acceptors (such as $V_{\rm B}$ or C$_{\rm N}$) in those samples.

Our calculations show that to avoid high compensation (and a potential collapse of mobility from charged-carrier scattering), only light donor doping should be utilized for AlN, and moderate doping should be used for $c$-BN. Moving to higher Si concentrations in these materials only creates additional Si$^+$ and Si$^-$ in equal concentrations. Again, we have assumed here that no other compensating species (such as unintentional impurities or native defects) are present in the semiconductor. If present, compensating acceptors (other than Si \textit{DX} centers) would linearly reduce the expected concentration of free electrons.

\section{Conclusions}
We have used DFT calculations of Si donor transition levels, along with temperature-dependent DFT calculations of band-edge positions, to examine how \textit{DX} centers cause compensation in Si-doped ultrawide-bandgap nitride semiconductors. In AlN, for which Si donors have a \textit{DX} level 271 meV below the conduction-band minimum, compensation by \textit{DX} centers is strong, and resulting free-carrier concentrations are strongly limited. This situation is improved by alloying with Ga, which moves the \textit{DX} level closer to the conduction band, and Si doping efficiency in AlGaN alloys is greatly {increased}. For $c$-BN, in which the Si \textit{DX} level is only 110 meV from the band edge, compensation by \textit{DX} centers still occurs, but free electron concentrations are significantly higher than in AlN.

\section{Acknowledgements}
This work was supported by the Office of Naval Research through the Naval Research Laboratory’s Basic Research Program. Calculations were supported in part by the DoD Major Shared Resource Center at AFRL.

%aipnum4-2.bst 2019-01-14 (MD) hand-edited version of apsrev4-1.bst

\end{document}